\documentclass[twocolumn,english,aps,prl,superscriptaddress,groupaddress,floatfix,graphics,graphicx]{revtex4-1}
\usepackage{lmodern}
\usepackage{lmodern}
\usepackage[T1]{fontenc}
\usepackage[latin9]{inputenc}
\usepackage{geometry}
\usepackage{color}
\usepackage{babel}
\usepackage{amsmath}
\usepackage{amssymb}
\usepackage{graphicx}
\usepackage[unicode=true,pdfusetitle,
 bookmarks=true,bookmarksnumbered=false,bookmarksopen=false,
 breaklinks=false,pdfborder={0 0 1},backref=false,colorlinks=false]
 {hyperref}
\hypersetup{
 colorlinks,linkcolor=blue,citecolor=blue,urlcolor=blue}
\usepackage{breakurl}

\makeatletter
\usepackage{babel}
\usepackage{babel}
\usepackage{babel}

\usepackage{blindtext}

\usepackage{dsfont}\usepackage{bbm}\IfFileExists{lmodern.sty}{\usepackage{lmodern}}{}
\usepackage{babel}

\makeatother

\begin{document}

\title{Optimal charge-to-spin conversion in graphene on transition metal
dichalcogenides }

\author{Manuel Offidani}

\affiliation{Department of Physics, University of York, York YO10 5DD, United
Kingdom}

\author{Mirco Milletarì }
\email{milletari@gmail.com}

\affiliation{Dipartimento di Matematica e Fisica, Università Roma Tre, 00146 Rome,
Italy}

\affiliation{Bioinformatics Institute, Agency for Science, Technology and Research
(A{*}STAR), Singapore 138671, Singapore}

\author{Roberto Raimondi}

\affiliation{Dipartimento di Matematica e Fisica, Università Roma Tre, 00146 Rome,
Italy}

\author{Aires Ferreira}
\email{aires.ferreira@york.ac.uk}

\affiliation{Department of Physics, University of York, York YO10 5DD, United
Kingdom}
\begin{abstract}
\textcolor{black}{When graphene is placed on a monolayer of semiconducting
transition metal dichalcogenide (TMD) its band structure develops
rich spin textures due to proximity spin\textendash orbital effects
with interfacial breaking of inversion symmetry. In this work, we
show that the characteristic spin winding of low-energy states in
graphene on TMD monolayer enables current-driven spin polarization,
a phenomenon known as the inverse spin galvanic effect (ISGE). By
introducing a proper figure of merit, we quantify the efficiency of
charge-to-spin conversion and show it is close to unity}\textcolor{black}{\emph{
}}\textcolor{black}{when the Fermi level approaches the spin minority
band. Remarkably, at high electronic density, even though sub-bands
with opposite spin helicities are occupied, the efficiency decays
only algebraically. The giant ISGE predicted for graphene on TMD monolayers
is robust against disorder and remains large at room temperature.} 
\end{abstract}
\maketitle
In the past decade, graphene has emerged as a strong contender for
next-generation spintronic devices due to its long spin diffusion
lengths at room temperature and gate tunable spin transport \cite{G_spintronics_review}.
However, the lack of a band gap and its weak spin\textendash orbit
coupling (SOC) pose major limitations for injection and control of
spin currents. In this regard, van der Waals heterostructures \cite{2D_materials_review}
built from stacks of graphene and other two-dimensional (2D) materials
hold great promise \cite{Spinorbitronics_review}. The widely tunable
electronic properties in vertically-stacked 2D crystals offer a practical
route to overcome the weaknesses of graphene \cite{2D_band_engineering}.
An ideal match to graphene are group-VI dichalcogenides $MX_{2}$
(e.g., $M=\textrm{Mo,\thinspace W}$; $X=\textrm{S,\thinspace Se}$).
The lack of inversion symmetry in TMD monolayers enable spin- and
valley-selective light absorption \cite{TMD}, thus providing all-optical
methods for manipulation of internal degrees of freedom \cite{Muniz_2015}.
The optical injection of spin currents across graphene\textendash TMD
interfaces has been recently reported \cite{G-TMD_Exp_Luo_17,G-TMD_Exp-Avsar_17},
following a theoretical proposal \cite{gmitra2015}. Furthermore,
electronic structure calculations show that spin\textendash orbital
effects in graphene on TMD are greatly enhanced \cite{gmitra2016,wang2015},
consistently with the SOC fingerprints in transport measurements \cite{avsar2014,wang2015,wang2016,Volkl16},
pointing to Rashba-Bychkov (RB) SOC in the range of 1\textendash 10
meV.

In this Letter, we show that the SOC enhancement in graphene on a
TMD monolayer allows for current-induced spin polarization, a relativistic
transport phenomenon commonly known as ISGE or the Edelstein effect
\cite{ISGE}. In the search for novel spintronic materials, the role
of the ISGE, together with its Onsager reciprocal\textemdash the spin-galvanic
effect\textemdash is gaining strength, with experimental reports in
spin-split 2D electron gases formed in Bi/Ag and $\text{LaAlO}_{3}\text{/\ensuremath{\text{SrTiO}_{3}}}$,
as well as in topological insulator (TI) $\alpha$-Sn thin films \cite{sanchez2013,lesne2016,sanchez2016}.
In addition, the enhancement of non-equilibrium spin polarization
has been proposed in ferromagnetic TMD and magnetically-doped TI/graphene
\cite{magnetic_ISGE}. The robust ISGE in nonmagnetic graphene/TMD
heterostructures predicted here promises unique advantages for low-power
charge-to-spin conversion (CSC), including the tuning of spin polarization
by a gate voltage. Moreover, owing to the Dirac character of interfacial
states in graphene on TMD monolayer, the ISGE shows striking similarities
to CSC mediated by ideal topologically protected surface states \cite{Schwab11},
allowing nearly optimal CSC. We quantify the CSC efficiency as function
of the scattering strength, and show it can be as great as $\approx30$\%
at room temperature (for typical spin\textendash orbit energy scale
smaller than $k_{B}T$).

\begin{figure*}[t]
\begin{centering}
\includegraphics[width=0.6\textwidth]{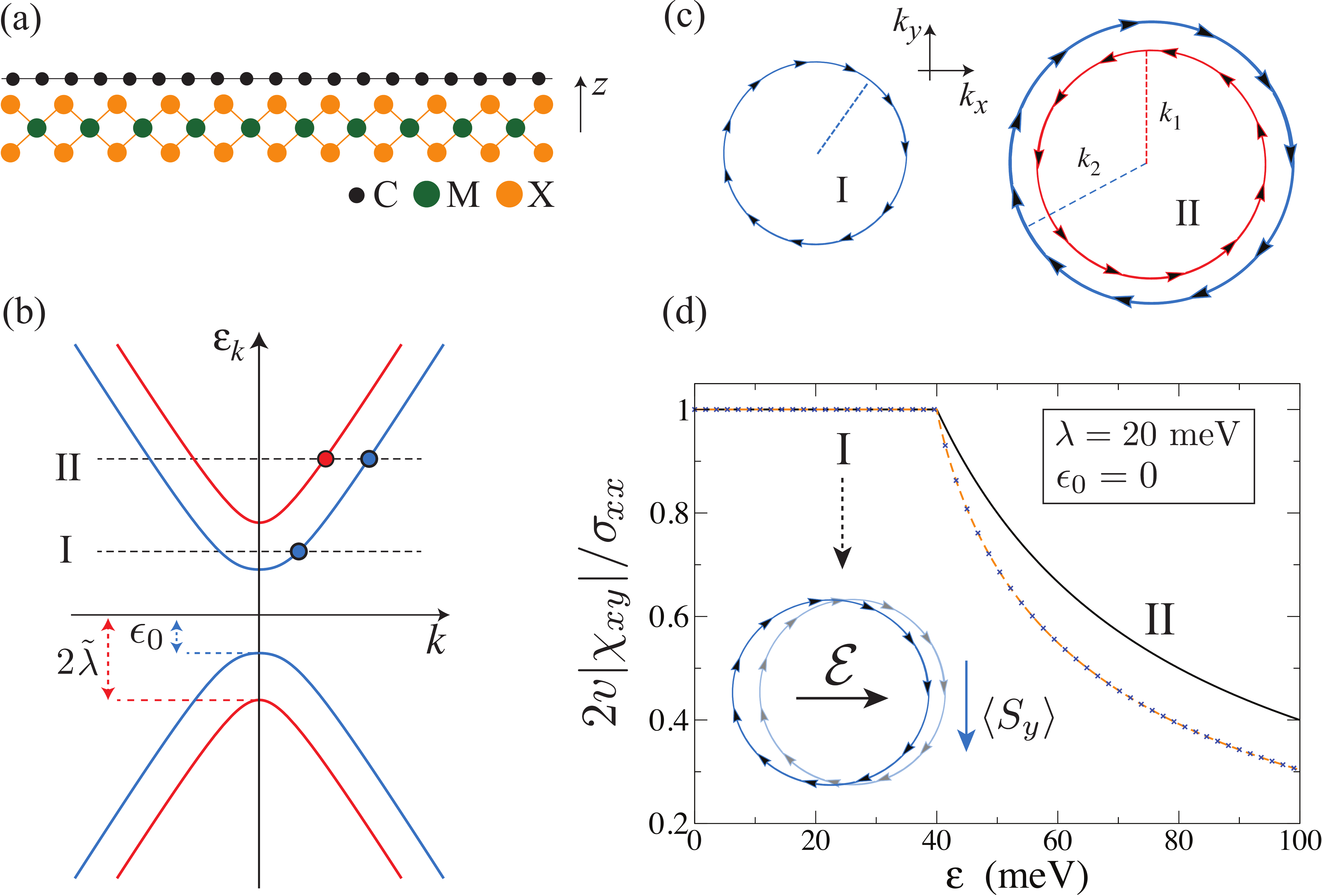} 
\par\end{centering}
\caption{\label{fig:01}(a) Graphene on a MX$_{2}$ monolayer. (b) Typical
band structure with spin-split bands with opposite spin helicity.
(c) Tangential winding of spin texture in regimes I and II. (d) Ratio
between the static spin\textendash charge susceptibility and charge
conductivity (in units of $2v$) {[}thick line (Born limit); dashed
line (strong scattering limit, $u_{0}\rightarrow\infty$){]}. }
\end{figure*}

\emph{The model.\textemdash }The electronic structure of graphene
on a TMD monolayer (G/TMD) is well described at low energies by a
Dirac model in two spatial dimensions \citep{wang2015,gmitra2016}
\begin{equation}
H_{0\mathbf{k}}=\tau_{z}\left[v\,\boldsymbol{\sigma}\cdot\mathbf{k}+\lambda\,\left(\boldsymbol{\sigma}\times\mathbf{s}\right)\cdot\hat{z}+\Delta\,\sigma_{z}+\lambda_{\textrm{sv}}\,s_{z}\right]\,,\label{eq:H0}
\end{equation}
where $\mathbf{k}=(k_{x},k_{y})$ is the 2D wavevector around a Dirac
point, $v$ is the Fermi velocity of massless Dirac electrons ($v\approx10^{6}$\,m/s)
and $\sigma_{i},s_{i},\tau_{i}\;(i=x,y,z)$ are Pauli matrices associated
with the sublattice, spin, and valley subspaces, respectively. The
momentum-independent terms in Eq.\,(\ref{eq:H0}) describe a RB effect
resulting from interfacial breaking of inversion symmetry ($\lambda$),
and staggered ($\Delta$) and spin\textendash valley ($\lambda_{\textrm{sv}}$)
interactions due to broken sublattice symmetry $C_{6v}\rightarrow C_{3v}$
{[}see Fig.~\ref{fig:01}\,(a){]}. The Dirac Hamiltonian $H_{0\mathbf{k}}$
contains all substrate-induced terms (to lowest order in $\mathbf{k}$)
that are compatible with time-reversal symmetry and the point group
$C_{3v}$ \cite{Kochan17}, except for a Kane\textendash Mele SOC
term ($\propto\sigma_{z}s_{z}$), which is too weak \citep{Hernando06,Fabian09}
to manifest in transport and can be safely neglected. The dispersion
relation associated with $H_{0\mathbf{k}}$ for each valley $\tau\equiv\tau_{z}=\pm1$
consists of two pairs of spin split Dirac bands (omitting $\hbar$)
\begin{equation}
\epsilon_{\tau\zeta}(k)=\pm\tau\sqrt{v^{2}k^{2}+\Delta_{\zeta}^{2}(k)}\,,\label{eq:dispersion_relation}
\end{equation}
where $k\equiv|\mathbf{k}|$, $\zeta=\pm1$ is the spin-helicity index
and 
\begin{align}
\Delta_{\zeta}^{2}(k) & =\Delta^{2}+\lambda_{\textrm{sv}}^{2}+2\lambda^{2}\nonumber \\
 & +2\zeta\sqrt{\left(\lambda^{2}-\Delta\lambda_{\textrm{sv}}\right)^{2}+v^{2}k^{2}\left(\lambda^{2}+\lambda_{\textrm{sv}}^{2}\right)}\,.\label{eq:delta_sq}
\end{align}
A typical spectrum is shown in Fig.\,\ref{fig:01}(b). The spin texture
associated with each band reads 
\begin{align}
\langle\mathbf{s}\rangle_{\alpha\mathbf{k}} & =-\zeta\,\varrho(k)\,(\hat{k}\times\hat{z})+m_{\alpha}^{z}(k)\,\hat{z}\,,\label{eq:spin_texture}
\end{align}
where $\alpha\equiv(\tau\zeta)$. The first term describes the spin
winding generated by the RB effect {[}Fig.~\ref{fig:01}(c){]} and
the second its out-of-plane tilting due to the broken sublattice symmetry.
The entanglement between spin and sublattice degrees of freedom generates
a nontrivial $k$ dependence in the spin texture. For example, in
the minimal model with only RB interaction, $\varrho(k)$ coincides
with the band velocity (in units of $v$), while $m_{\alpha}^{z}=0$,
i.e., the spin texture is fully in plane~\citep{rashba2009}. When
all interactions in Eq.\,(\ref{eq:H0}) are included, we find 
\begin{equation}
\varrho\left(k\right)=\frac{vk\lambda}{\sqrt{\left(\Delta\text{\ensuremath{\lambda}}_{\textrm{sv}}-\lambda^{2}\right)^{2}+v^{2}k^{2}\left(\lambda^{2}+\lambda_{\textrm{sv}}^{2}\right)}}\,.\label{eq:nu_texture_explicit}
\end{equation}
The breaking of sublattice symmetry modifies the spin texture, with
both valleys acquiring a spin polarization in the $\hat{z}$ direction,
consistently with first-principles studies \citep{gmitra2016}. The
explicit form of $m_{\alpha}^{z}(k)$ is too cumbersome to be presented.
Here, it is sufficient to note that $|m_{\alpha}^{z}(k=0)|=1$, with
$|m_{\alpha}^{z}(k)|$ decaying to zero away from the Dirac point\,\cite{SM}.
Finally, due to time-reversal symmetry the $\hat{z}$ polarizations
at inequivalent valleys are opposite. For energies within the Rashba
pseudo gap (RPG), that is, $\epsilon_{0}\equiv|\epsilon_{\tau-}(0)|<|\epsilon|<2\tilde{\lambda}\equiv|\epsilon_{\tau+}(0)|$,
the Fermi surface is simply connected. Hence, at low energies, the
electronic states have \emph{well-defined spin helicity} {[}Fig.\,\ref{fig:01}(b-c){]}.
This feature of G/TMD interfacial states is reminiscent of spin\textendash momentum
locking in topologically protected surface states \cite{Schwab11},
hinting at efficient CSC.

\emph{Semiclassical argument.\textemdash }The efficiency of CSC can
be demonstrated using a simple semiclassical argument. For ease of
notation, hereafter we employ natural units ($e\equiv1\equiv\hbar$).
Under a dc electric field, say $\vec{\mathcal{E}}=\mathcal{E}\,\hat{x}$,
the $\hat{y}$-polarization spin density in the steady state reads
$\langle S_{y}\rangle=\sum_{\alpha}\int(d\mathbf{k})\frac{1}{2}\langle s_{y}\rangle_{\alpha\mathbf{k}}\,\delta f_{\alpha\mathbf{k}}$,
where $\delta f_{\alpha\mathbf{k}}$ is the deviation of the quasiparticle
distribution function with respect to equilibrium and $(d\mathbf{k})\equiv d^{2}\mathbf{k}/4\pi{}^{2}$.
Owing to the tangential winding of the in-plane spin texture, only
the longitudinal component of the quasiparticle distribution function
$\delta f_{\alpha\mathbf{k}}^{\parallel}\equiv g_{\alpha}(k)\,\hat{k}\cdot\hat{k}_{x}$
contributes to the integral. At zero temperature, $g_{\alpha}(k)=\mp\mathcal{E}v_{\alpha k}\,\tau_{*\alpha k}\,\delta(\epsilon_{\alpha}(k)-\epsilon),$
where $v_{\alpha k}=\partial_{k}\epsilon_{\alpha}(k)$ is the band
velocity, $\tau_{*\alpha k}$ is the longitudinal transport time and
$\epsilon$ is the Fermi energy ($\mp$ for electron/holes). For energies
inside the RPG (regime I), one easily finds 
\begin{equation}
\langle S_{y}\rangle_{\textrm{I}}=\mp\frac{\mathcal{E}}{4\pi}\,\varrho(k_{F})\,k_{F}\tau_{*},\label{eq:S_y}
\end{equation}
where $k_{F}$ is the Fermi momentum and $\tau_{*}=\tau_{*(\tau-)k_{F}}$
(assumed valley-independent for simplicity). The charge current density,
$\langle J_{x}\rangle=-v\sum_{\alpha}\int(d\mathbf{k})\langle\tau_{z}\sigma_{x}\rangle_{\alpha\mathbf{k}}\,\delta f_{\alpha\mathbf{k}}$,
can be computed following identical steps. We obtain 
\begin{equation}
\langle J_{x}\rangle_{\textrm{I}}=\frac{\mathcal{E}}{2\pi}\,v_{F}\,k_{F}\,\tau_{*}\,,\label{eq:J_x}
\end{equation}
where $v_{F}=|v_{\tau-}(k_{F})|$. The implications of our results
are best illustrated by considering the minimal model, for which $\varrho(k_{F})=v_{F}/v$
and thus $\langle S_{y}\rangle_{\textrm{I}}=\mp\langle J_{x}\rangle_{\textrm{I}}/(2v)$.
Figure~\ref{fig:01}\,(d) shows the ratio of $\langle S_{y}\rangle/\langle J_{x}\rangle$
in the linear response regime computed according to the Kubo formula,
confirming the linear proportionality $\langle S_{y}\rangle_{\textrm{I}}\propto\langle J_{x}\rangle_{\textrm{I}}$.
The well-defined spin winding direction in regime I, responsible for
the semiclassical form of the non-equilibrium spin polarization {[}Eq.~(\ref{eq:S_y}){]},
automatically implies a large ISGE in the clean limit. Generally,
the CSC is optimal near the RPG edges, where $|\rho|$ is the largest
in regime I. In this energy range, the CSC \emph{is only limited by
the electronic mobility}, i.e., $|\langle S_{y}\rangle|_{\textrm{I}}\approx\langle J_{x}\rangle_{\textrm{I}}/(2v_{F})\propto(k_{F}\tau_{*})\,\mathcal{E}$.
\textcolor{black}{These considerations show that $|\varrho|\equiv|\langle S_{y}\rangle|/(2v_{F}\langle J_{x}\rangle)$
is the proper figure of merit i}n regime I. For models with $|\lambda_{\textrm{sv}}|\ll|\lambda|$,
the efficiency is \emph{nearly saturated} 
\begin{equation}
\max_{\epsilon\in\textrm{I};\lambda_{\textrm{sv}}=0}\,|\varrho(k(\epsilon))|=2\sqrt{2}/3\approx0.94,\label{eq:max_rho}
\end{equation}
\textcolor{black}{and is generally close to unity for not too large
spin\textendash valley coupling \cite{SM}.} In regime II, both spin
helicities $\zeta=\pm1$ contribute to the non-equilibrium spin density,
resulting into a decay of the CSC rate. Here,\textcolor{black}{{}
$|\varrho|$ is not a suitable figure of merit and an alternative
must be sought. As we show later, in this regime ($|\epsilon|>2\tilde{\lambda}$)}
the CSC efficiency exhibits an\textcolor{black}{{} algebraic decay
law, enabling a remarkably robust ISGE in typical experimental conditions. }

\begin{figure}[t]
\centering{}\includegraphics[width=0.7\columnwidth]{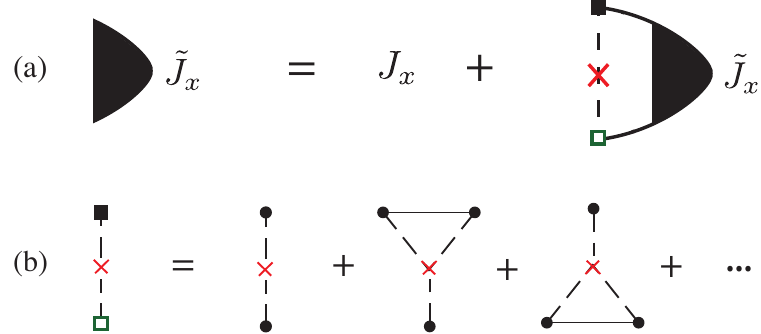}\caption{\label{fig:02}Diagrammatic expansion of the response function. (a)
Bethe-Salpeter equation for the charge current vertex in the $R$\textendash $A$
sector. (b) Skeleton expansion of the $T$-matrix ladder. Full (open)
square denotes a $T$ ($T^{\dagger}$) matrix insertion, while circles
represent electron\textendash impurity interaction vertices. The red
$\times$ stands for impurity density insertion ($n$).}
\end{figure}

\emph{Quantum treatment.\textemdash{}} To evaluate the full energy
dependence of the ISGE, we employ the self-consistent diagrammatic
approach developed by two of us in Ref.~\citep{MilletariFerreira2016}.
Despite the complexity of the Hamiltonian, Eq.\,(\ref{eq:H0}), one
can solve the Bethe\textendash Salpeter equations for the $T$-matrix
ladder. This provides accurate results in the regime $k_{F}v_{F}\tau_{*}\gg1$.
The zero-temperature spin density\textendash charge current response
function reads as 
\begin{align}
\chi_{yx}\left(\omega=0\right) & =\frac{1}{2\pi\Omega}\langle\text{Tr}\left[S_{y}\thinspace G^{+}\thinspace J_{x}\thinspace G^{-}\right]\rangle\,,\label{eq:spin-density-spin-current-RF}
\end{align}
where $G^{\pm}=(\epsilon-H\pm i0^{+})^{-1}$ is the Green's function
in the retarded/advanced sector of disordered G/TMD. Here, $\text{Tr}$
denotes the trace over internal and motional degrees of freedom, $\langle...\rangle$
stands for disorder average and $\Omega$ is the area. In the diagrammatic
approach, the disorder enters as a self-energy, $\Sigma^{a}$ ($a=\pm$),
``dressing'' the single-particle Green's functions, and as vertex
corrections in the electron\textendash hole propagator {[}Fig.~\ref{fig:02}\,(a){]}.
Since the response functions of interest are determined by the same
relaxation time, $\tau_{*}$, the CSC is expected to be little sensitive
to the disorder type as long as the latter is non magnetic. For practical
purposes, we use a model of short-range scalar impurities, $V(\mathbf{x})=u_{0}\sum_{i=1}^{N}\delta(\mathbf{x}-\mathbf{x}_{i})$,
where $\{\mathbf{x}_{i}=(x_{i},y_{i})\}$ are random impurity locations
and $u_{0}$ parametrizes their strength. This choice will enable
us to establish key analytical results across weak (Born) and strong
(unitary) scattering regimes.

We first evaluate Eq.\,(\ref{eq:spin-density-spin-current-RF}) for
models with fully in-plane spin texture, $\Delta,\lambda_{\textrm{sv}}=0$.
For ease of notation, we assume $\epsilon,\lambda>0$ in what follows.
The self-energy is given by $\Sigma^{a}=n\thinspace T^{a}$, where
$T^{a}=(u_{0}^{-1}\mathds{1}-g_{0}^{a})^{-1}$ and $n=N/\Omega$ is
the impurity areal density. Moreover, $g_{0}^{a}\equiv\int(d\mathbf{k})G_{0\mathbf{k}}^{a}$
and $G_{0\mathbf{k}}^{\pm}=(\epsilon-H_{0\mathbf{k}}\pm i0^{+})^{-1}$
is the bare Green's function. Neglecting the real part of $\Sigma^{a}$,
we have 
\begin{align}
\Sigma^{\pm} & =\mp in\left(\eta_{0}\gamma_{0}+\eta_{3}\,\gamma_{\textrm{KM}}+\eta_{r}\,\gamma_{r}\right)\,,\label{eq:self-en}
\end{align}
where $\gamma_{0}=\tau_{0}\sigma_{0}s_{0}$ (identity), $\gamma_{r}=\tau_{z}\left(\boldsymbol{\sigma}\times\mathbf{s}\right)\cdot\hat{z}$,
$\gamma_{\textrm{KM}}=\tau_{0}\sigma_{z}s_{z}$ and in the weak scattering
limit 
\begin{align}
\eta_{0} & =\frac{u_{0}^{2}}{8v^{2}}\left(\epsilon+\lambda\right),\,\eta_{3}=\frac{u_{0}^{2}}{8v^{2}}\lambda,\,\eta_{r}=-\frac{u_{0}^{2}}{16v^{2}}\epsilon\,,\label{eq:eta_coeff}
\end{align}
inside the RPG and $\eta_{0}=u_{0}^{2}\epsilon/4v^{2}$ and $\eta_{\textrm{KM}}=\eta_{r}=0$
for $\epsilon>2\lambda$ (see \cite{SM} for full $T$-matrix expressions).
The rich matrix structure in Eq.\,(\ref{eq:self-en}) stems from
the chiral (pseudo-spin) character of quasiparticles. In constrast,
in the 2D electron gas with RB spin\textendash orbit interaction,
the self energy due to spin-independent impurities is a scalar in
all regimes \cite{Schwab_02}. Next, we evaluate the disorder averaged
Green's function, $G_{\mathbf{k}}^{a}=[(G_{0\mathbf{k}}^{a})^{-1}-\Sigma^{a}]^{-1}$.
We define $\epsilon^{a}=\epsilon+i\,a\,n\,\eta_{0}$, $\lambda^{a}=\lambda-i\,a\,n\,\eta_{r}$,
and $m^{a}=i\,a\,n\,\eta_{3}$, which represent an energy shift, a
renormalized RB coupling and a random SOC gap, respectively. After
tedious but straightforward algebra we find
\begin{widetext}
\begin{align}
G_{\mathbf{k}}^{a}=-\left[\left(\epsilon^{a}\,L_{+}^{a}+\lambda^{a}L_{-}^{a}\right)\gamma_{0}+vL_{+}^{a}\tau_{z}\,\boldsymbol{\sigma}\cdot\mathbf{k}-\frac{1}{2}\left(\epsilon^{a}-m^{a}\right)L_{-}^{a}\gamma_{r}+\left(m^{a}L_{+}^{a}+\lambda^{a}L_{-}^{a}\right)\gamma_{\textrm{KM}}-vL_{-}^{a}\gamma_{v\mathbf{k}}+\Gamma_{\mathbf{k}}^{a}\right]\,\,,\label{eq:dis_av_propag}
\end{align}
\end{widetext}

where $L_{\pm}^{a}=(L_{1}^{a}\pm L_{2}^{a})/2$ with 
\begin{equation}
L_{1(2)}^{a}=[v^{2}k^{2}-(\epsilon^{a}-m^{a})(\epsilon^{a}+m^{a}\pm2\lambda^{a})]^{-1}\,,\label{eq:L12a}
\end{equation}
$\gamma_{v\mathbf{k}}=\tau_{0}\sigma_{0}(\hat{\mathbf{k}}\times\mathbf{s})\cdot\hat{z}$
and $\Gamma_{\mathbf{k}}^{a}$ is a $k_{i}$-quadratic term \cite{SM}.
The last step consists of evaluating the vertex corrections. The renormalized
charge current vertex satisfies the Bethe-Salpeter (BS) equation 
\begin{equation}
\tilde{J}_{x}=J_{x}+n\,\int(d\mathbf{k})\left\{ \,T^{+}\,\mathcal{G}_{\mathbf{\mathbf{k}}}^{+}\,\tilde{J}_{x}\,\mathcal{G}_{\mathbf{k}}^{-}\,T^{-}\,\right\} \,.\label{eq:BS_equation}
\end{equation}
The infinite set of non-crossing diagrams generated by the $T$-matrix
ladder describes incoherent multiple scattering events \emph{at all
orders} in the scattering strength $u_{0}$ {[}Fig.~\ref{fig:02}(b){]},
yielding an accurate description of spin\textendash orbit coupled
transport phenomena in the dilute regime \citep{MilletariFerreira2016}.
To solve Eq.\,(\ref{eq:BS_equation}), we decompose $\tilde{J}_{x}$
as $\tilde{J}_{x}=\tilde{J}_{x}^{\mu\nu\rho}\tau_{\mu}\sigma_{\nu}s_{\rho}$,
where the repeated indices $\mu,\nu,\rho\equiv\{0,i\}$ are summed
over. The number of nonzero components $\tilde{J}_{x}^{\mu\nu\rho}$
is constrained to only four by the symmetries of G/TMD \cite{note_symmetries}:
$(\mu,\nu,\rho)=\{(0,0,y),(z,x,0),(0,z,x),(z,y,z)\}$. Exploring the
properties of the Clifford algebra, one can show that the nonzero
vertex components have a one-to-one correspondence to their associated
non-equilibrium response functions \cite{milletari2017}. This allows
us to express $\chi_{yx}$ in terms of the spin density component
\emph{only}, $\tilde{J}_{x}^{\textrm{s}}\equiv\tilde{J}_{x}^{00y}$,
i.e., $\chi_{yx}=F_{s}(u_{0})\,\tilde{J}_{x}^{\textrm{s}}$, where
\begin{align}
\tilde{J}_{x}^{\textrm{s}} & =-\frac{v}{\epsilon}\,\frac{\epsilon^{2}\left(\epsilon+2\lambda\right)+\theta(\epsilon-2\lambda)\left(8\lambda^{3}-\epsilon^{3}\right)}{\epsilon^{2}+4\lambda^{2}}+\varepsilon_{\Lambda}.\label{eq:vertex_xs}
\end{align}
Here, $\theta$ is the Heaviside step function and $\varepsilon_{\Lambda}$
is a weak correction logarithmic in the ultraviolet cutoff $\Lambda$
set by the inverse of the lattice scale \citep{ferreira2011}. Finally,
$F_{s}(u_{0})$ is a complicated function, which in the Gaussian and
unitary scattering limits takes the form 
\begin{equation}
F_{s}(u_{0})=\frac{1}{2\pi n}\times\begin{cases}
\frac{4}{u_{0}^{2}} & ,|g_{0}^{+}u_{0}|\ll1\\
\left(\frac{\epsilon}{2\pi v^{2}}\log\left|\frac{\Lambda^{2}}{\epsilon\sqrt{\epsilon^{2}-4\lambda^{2}}}\right|\right)^{2} & ,|u_{0}|\rightarrow\infty
\end{cases},\label{eq:F_}
\end{equation}
respectively. Analogously, we can determine the expression for the
charge conductivity $\sigma_{xx}=F_{c}(u_{0})\tilde{J}_{x}^{c}$,
with $\tilde{J}_{x}^{\textrm{c}}\equiv\tilde{J}_{x}^{zx0}$ \cite{SM}.
The CSC rate can now be determined 
\begin{align}
-\frac{2v\chi_{yx}}{\sigma_{xx}} & =\theta(2\lambda-\epsilon)+\frac{2\lambda}{\epsilon}g(u_{0},\epsilon)\,\theta(\epsilon-2\lambda)\,,\label{eq:efficiency_Rashba_Kubo}
\end{align}
\textcolor{black}{where $g(u_{0},\epsilon=2\lambda)=1$ and deviates
only slightly from this value when $u_{0}$ is large and for $\epsilon>2\lambda$
{[}see Fig.\,\ref{fig:01}(d){]}. The central result Eq.\,(\ref{eq:efficiency_Rashba_Kubo})
puts our earlier semiclassical argument on firm grounds, and shows
that the CSC is little affected by the disorder strenght outside the
RPG.}

\begin{figure}[t]
\centering{}\textcolor{black}{\includegraphics[width=0.95\columnwidth]{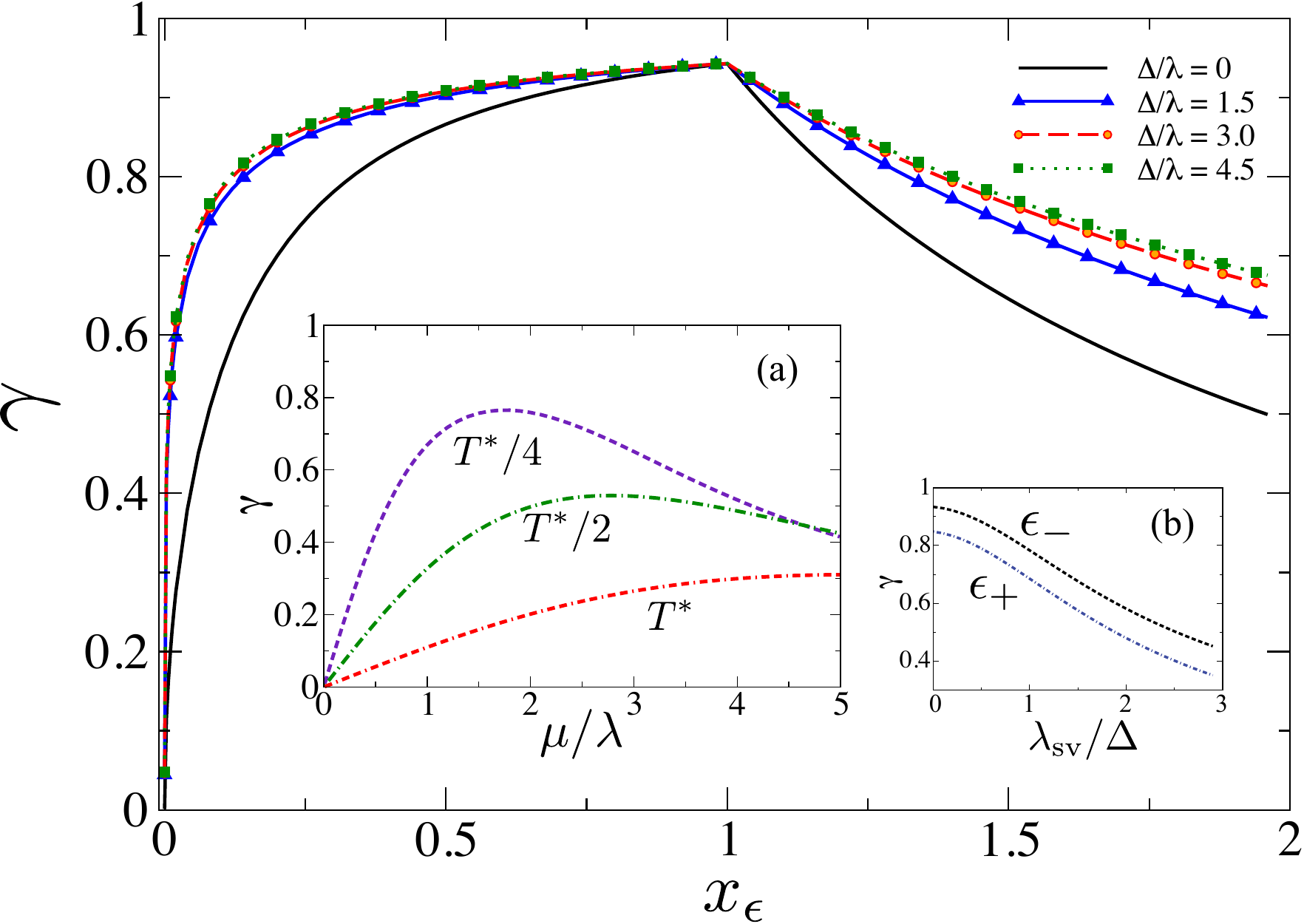}\caption{\label{fig:03}\textcolor{black}{Main figure: Fermi energy dependence
of the ISGE efficiency at selected values of $\Delta$ for $\lambda=15$~meV
and $\lambda_{\textrm{sv}}=0$. The $x$-axis is rescaled as $x_{\epsilon}=|\epsilon-\epsilon_{0}|/\left|2\tilde{\lambda}-\epsilon_{0}\right|$
for clarity. Insets: (a) $\gamma$ as function of chemical potential
($\mu$) at selected temperatures for a prototypical heterostructure
with $\lambda=10$ meV, $\lambda_{\textrm{sv}}=\Delta=0$ (Ref.\,\citep{wang2016});
$k_{B}T^{*}=25\,$meV (room temperature). (b) Variation of $\gamma$
with $\lambda_{\textrm{sv}}$ for a Fermi energy slightly below (above)
the RPG's edge {[}$\epsilon_{\pm}=2\tilde{\lambda}\times(1.00\pm0.05)${]}
for $\Delta=\lambda/2$ and $\lambda=15$~meV. All calculations performed
in the weak scattering limit.}}
} 
\end{figure}

\emph{Discussions.\textemdash{}} In realistic G/TMD heterostructures,
$\Delta$ and $\lambda_{\textrm{sv}}$ can be comparable to the RB
coupling \citep{gmitra2016}, leading to major modifications in the
band structure. Nevertheless, a thorough analysis, summarized in Fig.~\ref{fig:03},
shows that the ISGE remains robust. For instance, for $|\lambda_{\textrm{sv}}|\ll\lambda,|\Delta|$,
the $k$ dependence of the in-plane spin texture is virtually unaffected
{[}Eq.\,(\ref{eq:nu_texture_explicit}){]}. Thus, according to the
semiclassical results the CSC efficiency should be high at the RPG
edge. This is confirmed by a numerical inversion of the Bethe-Salpeter
equations in the full model. The figure of merit $\gamma$ plotted
in Fig.~\ref{fig:03} reaches its predicted optimal value {[}Eq.\,(\ref{eq:max_rho}){]}.
When the spin\textendash valley coupling is significant, the in-plane
spin texture shrinks, however the CSC efficiency remains sizeable
{[}Fig.~\ref{fig:03}(b){]}. Outside the RPG, the definition of efficiency
$\gamma$ is complicated due to the coexistence of counter-rotating
spins. To analyze this regime, we employ a heuristic definition satisfying:
(i) $0\le\gamma\le1$ for all parameters, (ii) $\gamma$ decays for
$\epsilon\gg2\tilde{\lambda}$ due to collapsing of spin-split Fermi
rings and (iii) $\gamma$ is continuous across the RPG. Since the
band velocity saturates quickly to its upper bound ($=v$), we use
its value at the RPG edge as representative for the regime II, which
lead us to the following definition 
\begin{equation}
\gamma=\frac{2|\chi_{yx}|}{\sigma_{xx}}\times\begin{cases}
v_{F}(\epsilon) & ,\,\epsilon<2\tilde{\lambda}\\
v_{F}(2\tilde{\lambda}) & ,\,\epsilon\ge2\tilde{\lambda}
\end{cases}\,.\label{eq:fig_merit}
\end{equation}
where $v_{F}(\epsilon)\equiv|v_{\tau-}(k(\epsilon))|$. Consistently
with the rate derived for the minimal model {[}Eq.\,(\ref{eq:efficiency_Rashba_Kubo}){]},
the asymptotic behavior of $\gamma$ is of power-law type, and thus
the CSC remains robust in the accessible range of electronic densities.
A relevant question is how much efficiency is lost due to thermal
fluctuations. Figure\,\ref{fig:03}(b) shows the CSC figure of merit
at selected temperatures in the weak scattering limit (see Ref.\,\cite{SM}
for methods). Since the $T=0$ ratio decays slowly in regime II, the
smearing caused by thermal activation is ineffective, allowing a giant
ISGE at room temperature, e.g., $\gamma_{\textrm{room}}\approx0.3$
for a chemical potential $\mu\approx5\lambda\approx50$ meV. We finally
comment on the rippling of the graphene surface and imperfections
causing local variations in the RPG \cite{Ripples}. Inhomogeneities
in the spin\textendash orbit energy scales are expected to be small
in samples with strong interfacial effect \cite{Yang_17}. As long
as $|\lambda(\mathbf{x})-\lambda|\ll\lambda$, the random spin\textendash orbit
field acts merely as an additional source of scattering \cite{SM},
which according to our findings would not affect the ISGE efficiency.

\textcolor{black}{In conclusion, we have presented a rigorous theory
of inverse spin galvanic effect for graphene on transition metal dichalcogenide
monolayers. We introduced a figure of merit for charge-to-spin conversion
and show it attains values close to unity at the minority spin band
edge. The effect is robust against nonmagnetic disorder and remains
large at room temperature. }The current-driven spin polarization is
only limited by the electronic mobility, and thus it is expected to
achieve unprecedentedly large values in ultra-clean samples\textcolor{black}{.
Our results are also relevant for} group-IV honeycomb layers\,\,\cite{Honeycomb_Layers},
which are described by similar Dirac models.

The codes used for numerical analyses are available from the Figshare
database, under the Ref\,\cite{figshare}.

M.M. thanks C. Verma for his hospitality at the Bioinfomatics Institute
in Singapore. A.F. gratefully acknowledges the financial support from
the Royal Society (U.K.) through a Royal Society University Research
Fellowship. R.R. acknowledges the hospitality of CA2DM at NUS under
grant R-723-000-009-281 (GL 769105). M.O. and A.F. acknowledge funding
from EPSRC (Grant Ref: EP/N004817/1). 

\newpage{}

\onecolumngrid

\section*{{\large{}Supplementary information for ``OPTIMAL CHARGE-TO-SPIN
CONVERSION IN GRAPHENE ON TRANSITION METAL DICHALCOGENIDES''}}

In this Supplementary Information we provide additional details on
the Dirac-Rashba model and the semiclassical theory at large electronic
density. We also provide the explicit form of the renormalized charge
current vertex for the minimal model (i.e., $\Delta=\lambda_{\textrm{sv}}=0$,
$\lambda\neq0$), as well as additional details on the finite temperature
calculation and the impact of random fluctuations in the spin\textendash orbit
energy scale.

\section{DETAILS ON THE MODEL}

\subsection{Spectrum }

\begin{figure}[h]
\includegraphics[scale=0.5]{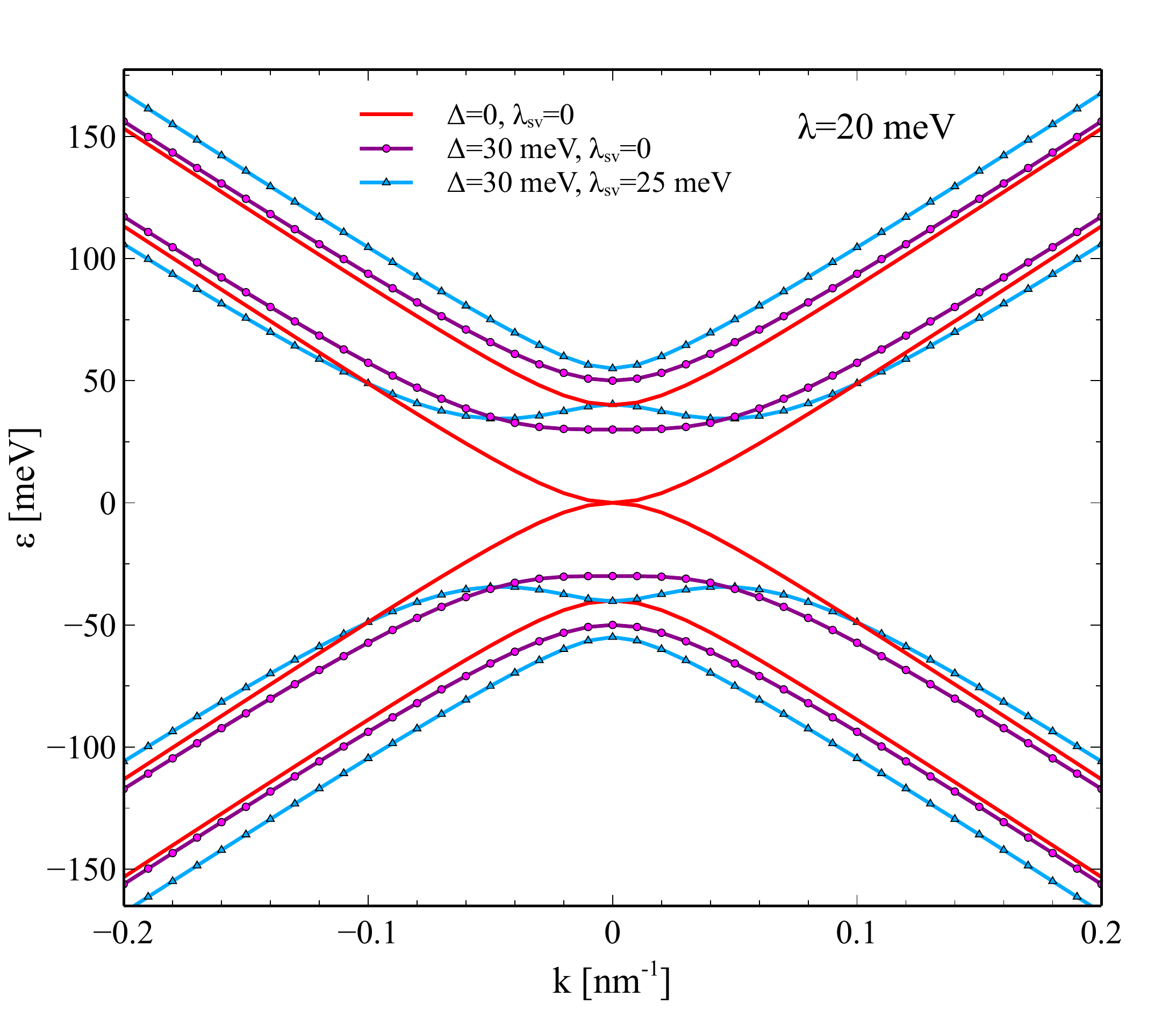}

\caption{Spectrum of the full model. For $\Delta=0,\lambda_{\text{sv}}=0$
(solid lines) the spectrum displays a Rashba pseudogap, with a simply
connected Fermi surface at low energies and no band gap. A finite
staggered potential, $\Delta=30$ meV, opens a gap (circles), which
in combination with $\lambda_{\text{sv}}=25$ meV creates a ``Mexican
hat'' between $\epsilon_{m}$ and $\delta/2$ (triangles). \label{fig:Spectrum-of-the}}
\end{figure}

The effective Hamiltonian of graphene on TMD monolayer can be written
as \cite{gmitra2016,wang2015} 
\begin{equation}
H_{0\mathbf{k}}=\tau_{z}\left[v\,\boldsymbol{\sigma}\cdot\mathbf{k}+\lambda\,\left(\boldsymbol{\sigma}\times\mathbf{s}\right)\cdot\hat{z}+\Delta\,\sigma_{z}+\lambda_{\textrm{sv}}\,s_{z}\right]\,,\label{eq:Ham}
\end{equation}
where $\boldsymbol{\sigma},\mathbf{s}$ are Pauli matrices and we
have used the representation for the 4-component spinors at each valley
($\tau_{z}=\pm1$):
\begin{equation}
\psi_{\tau_{z}=\pm}=(\psi_{\pm,a(b)}^{\uparrow},\psi_{\pm,a(b)}^{\downarrow},\psi_{\pm,b(a)}^{\uparrow},\psi_{\pm,b(a)}^{\downarrow})^{\textrm{t}}\,.\label{eq:representation}
\end{equation}
In the above, $a(b)$ are graphene sublattice indexes, and $\uparrow,\downarrow$
denote the spin projection. The respective eigenvalues are given in
Eqs.\ (2)-(3) of main text. The Rashba pseudo-gap at $k=0$ (see
Fig.\,1, main text) is easily computed as 
\begin{equation}
2\tilde{\lambda}=\min\{|\Delta+\lambda_{\textrm{sv}}|,\,\sqrt{4\lambda^{2}+(\Delta-\lambda_{\textrm{sv}})^{2}}\},
\end{equation}
while the bottom of the spin majority conduction band is 
\begin{equation}
\epsilon_{m}=\frac{|\lambda(\Delta+\lambda_{\text{sv}})|}{\sqrt{\lambda^{2}+\lambda_{\text{sv}}^{2}}}\,.
\end{equation}
For energies $\epsilon_{m}<\epsilon<2\tilde{\lambda}$ the spectrum
develops a small ``Mexican hat'' feature \cite{gmitra2016}. In
Fig.\ \ref{fig:Spectrum-of-the} we show the evolution of the spectrum
for finite Rashba effect as one turns on the proximity couplings $\Delta,\lambda_{\text{sv}}.$
We note that the energy spectrum is gapless in the following particular
cases: (i) $\lambda=0\text{ and }|\lambda_{\text{sv}}|>|\Delta|$
and (ii) $\lambda_{\text{sv}}=-\Delta$.

\begin{figure}
\includegraphics[scale=0.5]{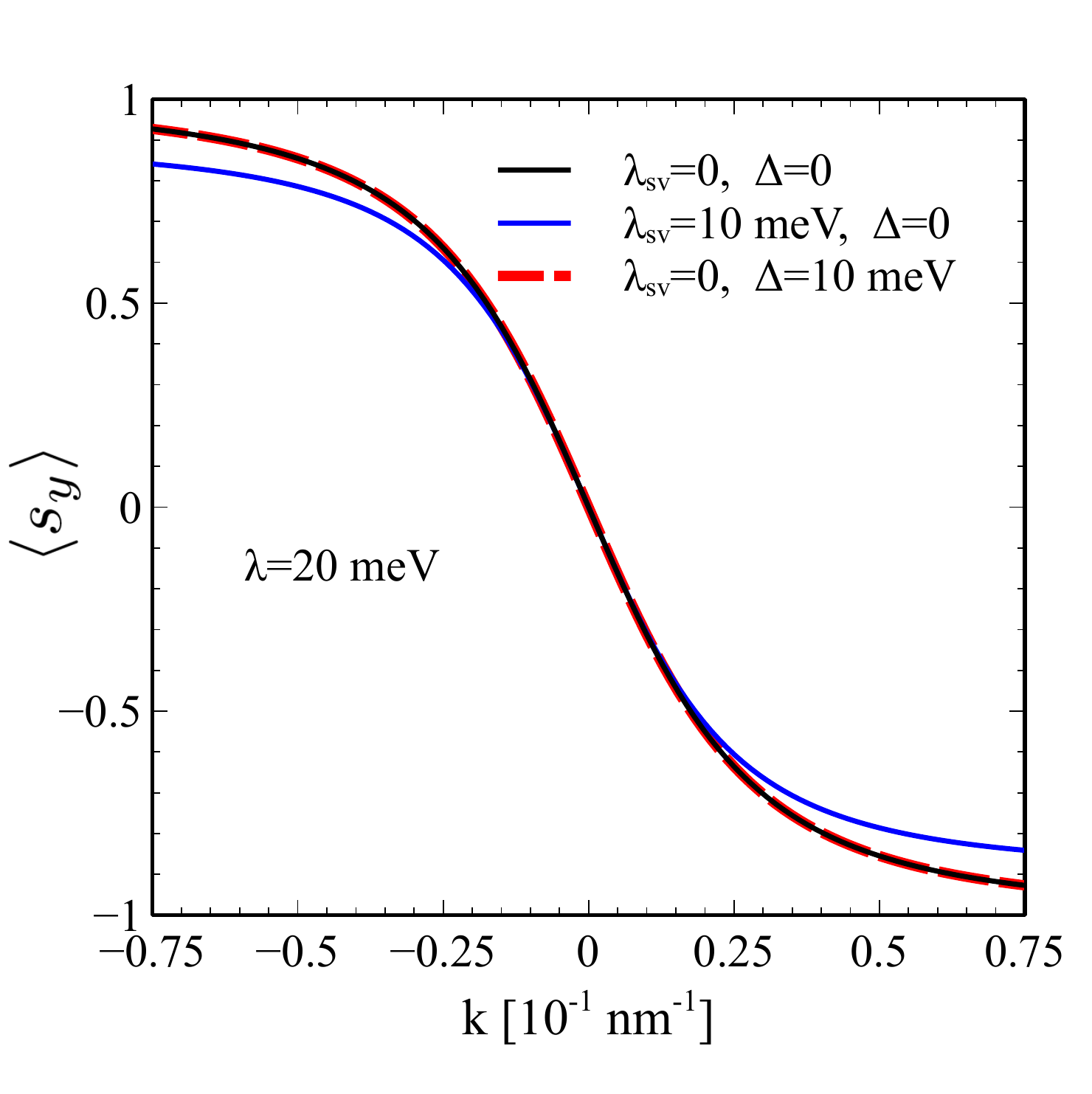}\includegraphics[scale=0.51]{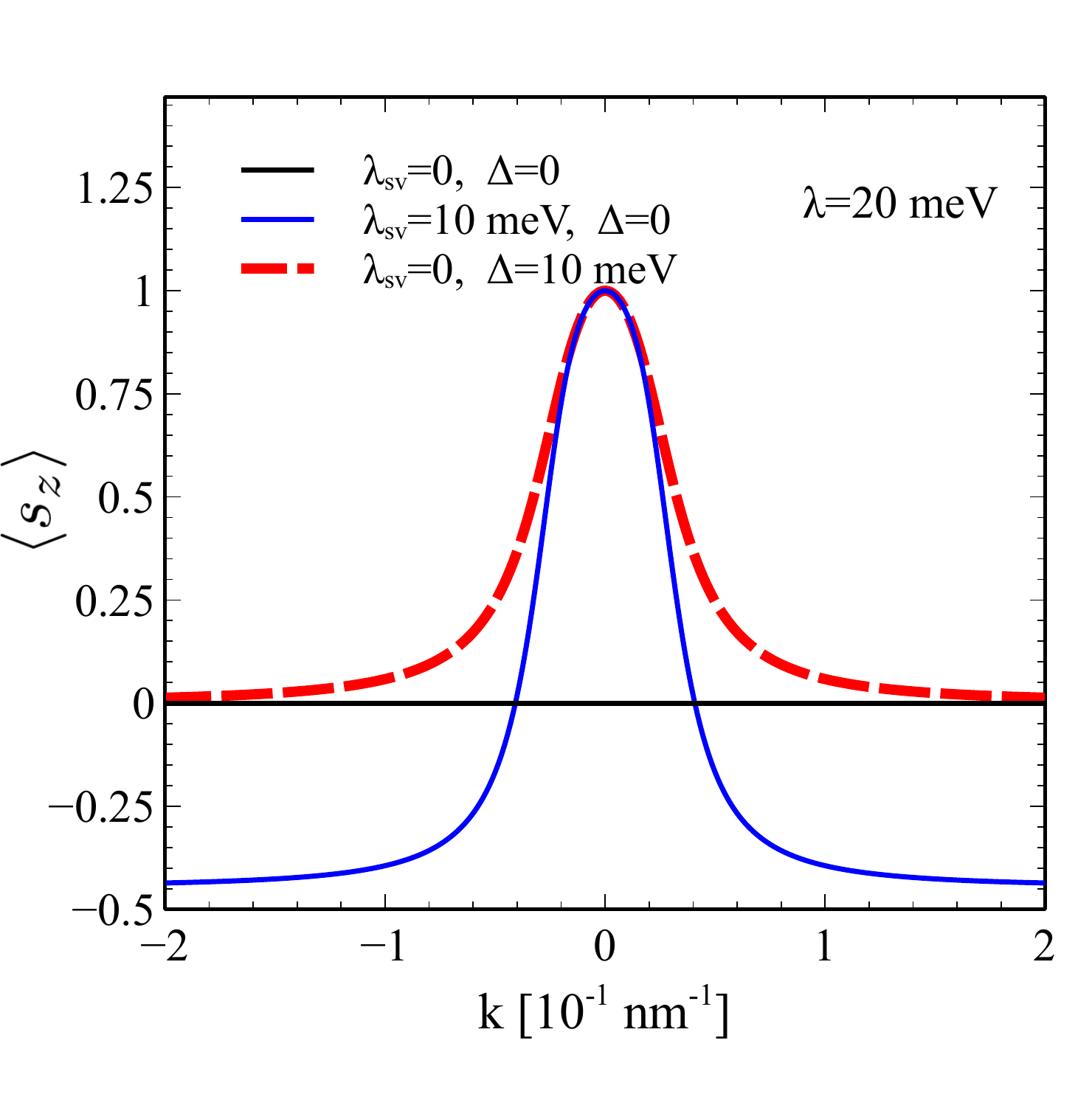}\caption{Spin texture for electronic states along $\hat{k}_{x}$ in the spin-majority
conduction band. In the pure Rashba model the spins lie entirely in
the plane, $\langle s_{z}\rangle=0$. When $\lambda_{\text{sv}}$
or $\Delta$ are switched on, electrons develop a $\langle s_{z}\rangle$
component. Interestingly, for $\Delta\protect\neq0$ and $\lambda_{\textrm{sv}}=0$
the latter does not imply a decrease of $\langle s_{y}\rangle$ (see
left panel).\label{fig:-components-for}}
\end{figure}

In the minimal model ($\Delta,\lambda_{\textrm{sv}}=0$) the spin
texture is entirely in-plane, due to the Rashba spin-momentum locking.
The additional proximity-induced couplings in Eq.\,\eqref{eq:Ham}
favor the establishment of a finite $s_{z}$-component. In Fig.\,\ref{fig:-components-for},
we show the spin texture of the electron spin-majority band for a
number of representative cases. 

\subsection{Semiclassical interpretation of the large energy behavior of the
spin\textendash charge response function}

We demonstrate how the asymptotic scaling of the ISGE efficiency reported
in the main text {[}viz., Eq.\,(17){]} can be understood within a
simple semiclassical picture. For simplicity we study the pure-Rashba
model, where $m_{\zeta}^{z}(k)=0$. The argument can be easily generalized
for other cases. Neglecting interband transitions, the spin-$y$ linear
response to an electric field applied along $\hat{x}$ axis is given
by {[}cf. Eq. (6) of the main text{]} 
\begin{equation}
\chi_{yx}=\frac{1}{4\pi}\int\frac{d\theta_{\boldsymbol{k}}}{2\pi}\sum_{\zeta=\pm1}\langle s_{y}\rangle_{\zeta}\cos\theta_{\boldsymbol{k}}\,k_{\zeta}(\epsilon)\tau_{*\zeta}(\epsilon)\label{eq:chi-1}
\end{equation}
where $k_{\zeta}(\epsilon)=v^{-1}\sqrt{\epsilon(\epsilon-\zeta\lambda)}$
are the Fermi radii. Substituting the expression for the equilibrium
spin texture {[}Eq.\,(5); main text{]}, we find, for large $\epsilon$
\begin{align}
\chi_{yx} & \simeq-\frac{1}{8\pi}\sum_{\zeta=\pm1}\zeta\,\frac{\epsilon}{v}\left(1-\zeta\frac{\lambda}{2\epsilon}+...\right)\tau_{*\zeta}(\epsilon)\,.\label{eq:aux}
\end{align}
Using the form of the momentum relaxation time in the Gaussian and
unitary limits, we find, respectively 
\begin{align}
\tau_{*\zeta}(\epsilon) & =A/\epsilon\quad\longrightarrow\quad\chi_{yx}=\frac{A}{8\pi v}\frac{\lambda}{\epsilon}\,,\label{eq:gauss}\\
\tau_{*\zeta}(\epsilon) & =A^{\prime}\epsilon\quad\longrightarrow\quad\chi_{yx}=\frac{A^{\prime}}{8\pi v}\lambda\,\epsilon\,.\label{eq:un}
\end{align}
While the collapsing of the Fermi rings, $k_{\zeta}(\epsilon)\to k_{-\zeta}(\epsilon)$
as $\epsilon\gg2|\lambda|$, tends to diminish the out-of-equilibrium
spin polarization, the latter can still be finite depending on the
asymptotic behavior of $\tau_{*}(\epsilon)$. In the unitary limit,
one has $vk_{\zeta}(\epsilon)\tau_{*\zeta}(\epsilon)\propto\epsilon^{2}$,
resulting in a monotonically increasing spin\textendash charge response
function. However, the ratio between the spin\textendash charge response
function and the charge conductivity is always $\chi_{yx}/\sigma_{xx}\propto\epsilon^{-1}$,
as shown in the main text. While in Eqs.\,(\ref{eq:gauss})-(\ref{eq:un})
we have neglected the role of scattering between states with different
spin helicity, the latter processes are included in the quantum-mechanical
treatment in the main text. Given the agreement of Eqs.\,(\ref{eq:gauss})-(\ref{eq:un})
with Eq.\,(16) of the main text, we conclude that their inclusion
will not affect the above semiclassical picture.

\section{DETAILS ON THE DIAGRAMMATIC CALCULATION}

\subsection{Disorder averaged propagators}

We provide the full form of the disorder averaged propagator in the
pure-Rashba model. Denoting with $a=\pm$, respectively, the retarded
and advanced sector of the theory, we obtain 
\begin{equation}
G_{\mathbf{k}}^{a}=-\left[\left(\epsilon^{a}L_{+}^{a}+\lambda^{a}L_{-}^{a}\right)\gamma_{0}+vL_{+}^{a}\tau_{z}\boldsymbol{\sigma}\cdot\mathbf{k}-\frac{1}{2}\left(\epsilon^{a}-m^{a}\right)L_{-}^{a}\gamma_{r}+\left(m^{a}L_{+}^{a}+\lambda^{a}L_{-}^{a}\right)\gamma_{\textrm{KM}}-vL_{-}^{a}\tau_{0}\sigma_{0}(\hat{\mathbf{k}}\times\mathbf{s})\cdot\hat{z}+\Gamma_{\mathbf{k}}^{a}\right]\,,
\end{equation}
where 
\begin{equation}
\epsilon^{a}=\epsilon+i\,a\,n\,\eta_{0}\,,\;\lambda^{a}=\lambda-i\,a\,n\,\eta_{r}\,,\;m^{a}=-i\,a\,n\,\eta_{3}\,,\label{eq:aux-1}
\end{equation}
\begin{equation}
L_{\pm}^{a}=(L_{1}^{a}\pm L_{2}^{a})/2\,,\quad L_{1(2)}^{a}=[v^{2}k^{2}-(\epsilon^{a}-m^{a})(\epsilon^{a}+m^{a}\pm2\lambda^{a})]^{-1},\label{eq:L12a-1}
\end{equation}
and 
\begin{align}
\Gamma_{\mathbf{k}}^{a} & =\left(\lambda^{a}L_{+}^{a}\tau_{0}+\frac{\left(\epsilon^{a}+m^{a}\right)}{2}L_{-}^{a}\tau_{z}\right)\left[\sin2\phi_{\boldsymbol{k}}\left(\sigma_{x}s_{x}-\sigma_{y}s_{y}\right)-\cos2\phi_{\boldsymbol{k}}\tau_{z}\left(\sigma_{x}s_{y}+\sigma_{y}s_{x}\right)\right]\,.
\end{align}

\subsection{$T$-matrix calculation}

We report the full form of the imaginary part of the self-energy in
the $T$-matrix approximation:
\begin{equation}
\Sigma^{\pm}=\mp in\left(\eta_{0}\gamma_{0}+\eta_{3}\,\gamma_{\textrm{KM}}+\eta_{r}\,\gamma_{r}\right),\label{eq:Self-En}
\end{equation}
\begin{align}
\eta_{0},\eta_{3} & =\frac{u_{0}}{2}\,\text{Im}\,\left[\frac{1}{1-u_{0}\left(g_{0,0}^{+}+g_{0,\text{KM}}^{+}\right)}\pm\frac{1-u_{0}\left(g_{0,0}^{+}-g_{0,\text{KM}}^{+}\right)}{\left[1-u_{0}\left(g_{0,0}^{+}-g_{0,\text{KM}}^{+}\right)\right]^{2}-\left(2u_{0}\,g_{0,r}^{+}\right)^{2}}\right]\,,\label{eq:eta0,eta3}\\
\eta_{r} & =\text{Im}\,\left[\frac{u_{0}g_{0,r}^{+}}{\left[1-u_{0}\left(g_{0,0}^{+}-g_{0,\text{KM}}^{+}\right)\right]^{2}-\left(2u_{0}\,g_{0,r}^{+}\right)^{2}}\right]\,,\label{eq:etaR}
\end{align}
with $g_{0,i}^{+}=\frac{1}{8}\,\text{Tr}[g_{0}^{+}\gamma_{i}]$ and
$\gamma_{i}=\left\{ \gamma_{0},\gamma_{\text{KM}},\frac{1}{2}\gamma_{r}\right\} $
as defined in the main text. The real part of the self-energy (omitted
for simplicity) leads to a renormalization of the Fermi energy and
of $\lambda$ as well as a random mass term of the Kane-Mele type
\cite{MilletariFerreira2016}. The Fermi energy renormalization contains
a logarithmic divergence, which can be taken into account by wave
function renormalization and leads to a renormalization of the Fermi
velocity \cite{Vozmediano_11}. 

\subsection{Full form of the renormalized charge vertex}

\begin{figure}
\includegraphics[scale=0.5]{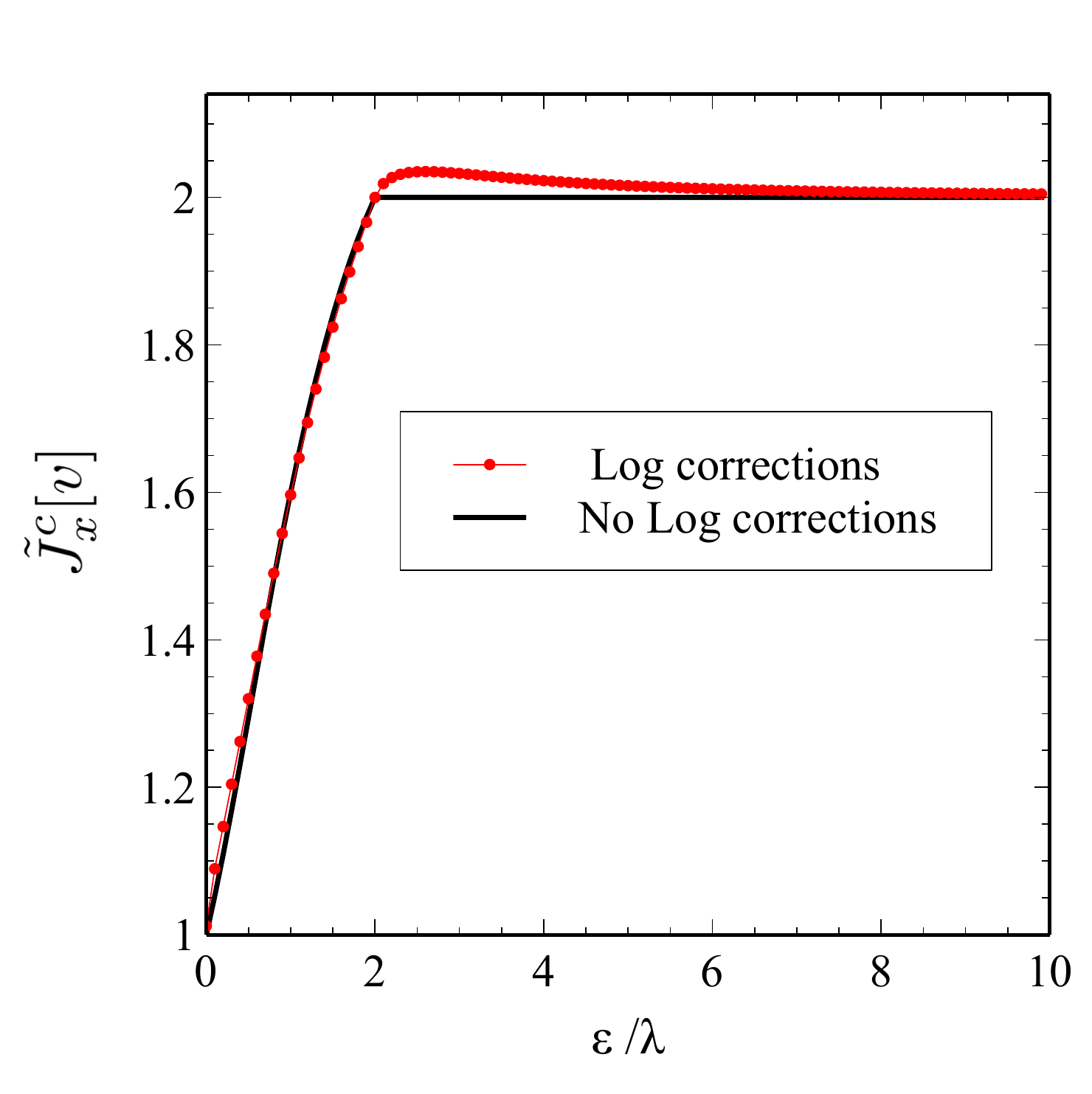}\includegraphics[scale=0.5]{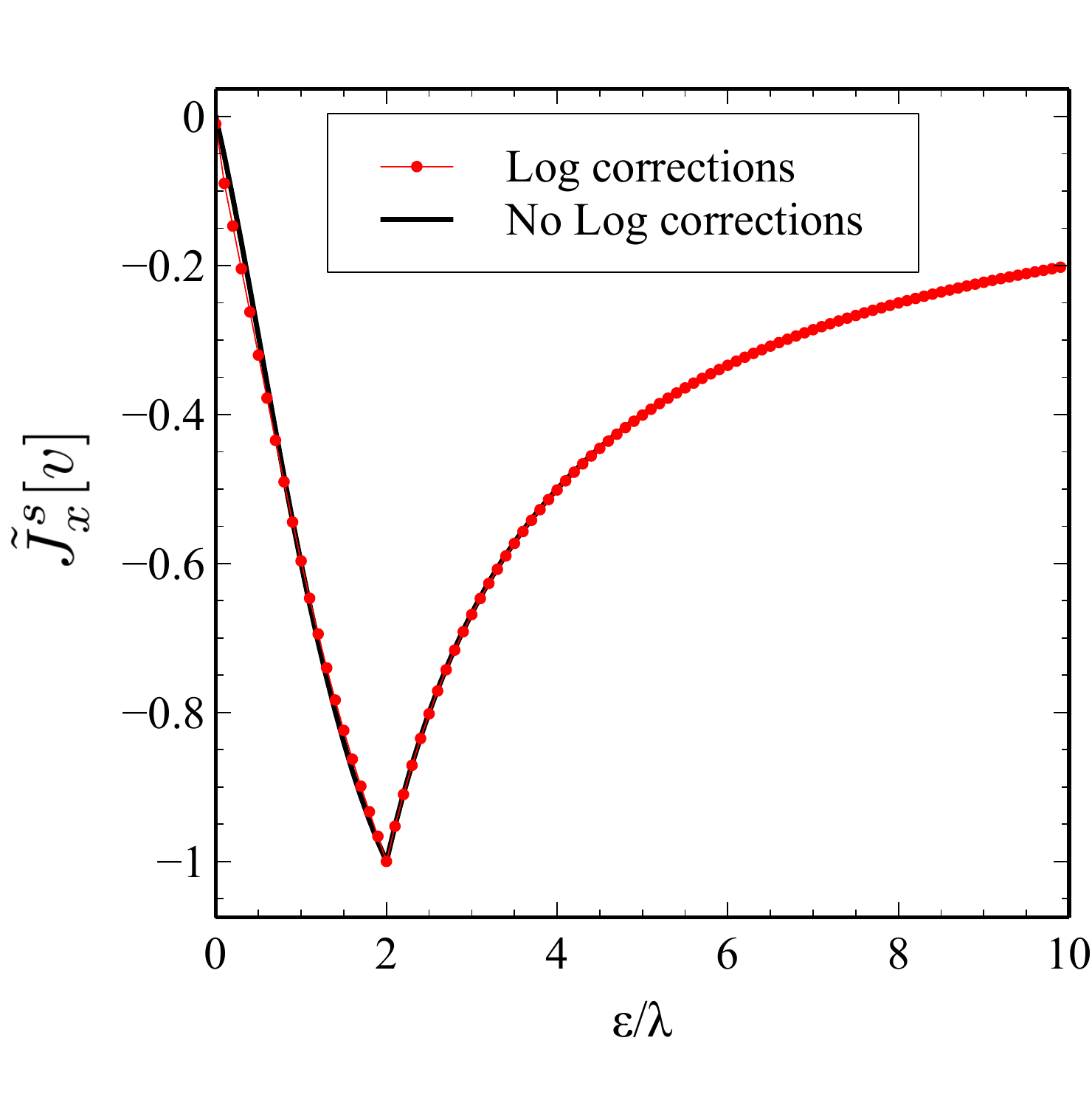}\caption{The charge $\tilde{J}_{x}^{c}$ and spin density $\tilde{J}_{x}^{s}$
component of the renormalized charge vertex for $\lambda=25$~meV.
Corrections logarithmic in the cutoff $\Lambda=10$\,eV are shown
to be small.\label{fig:The-charge-}}
\end{figure}

We first define the general structure of the renormalized charge current
vertex in the minimal model as 
\begin{equation}
\left\{ \tilde{J}_{x}^{c},\,\tilde{J}_{x}^{s}\,,\tilde{J}_{x}^{sh},\tilde{J}_{x}^{m}\right\} =\frac{1}{8}\text{Tr}\left[\tilde{J}_{x}\left\{ \tau_{z}\sigma_{x},s_{y},\tau_{z}\sigma_{y}s_{z},\sigma_{z}s_{x}\right\} \right].\label{eq:struct}
\end{equation}
Below we provide the explicit form of the components to lowest order
in the impurity density $n$ for the weak scattering limit. For simplicity,
we assume $\lambda>0$. Outside the Rashba pseudogap, $\epsilon>2\lambda$,
we obtain

\begin{align}
\tilde{J}_{x}^{c} & =2v+\mathcal{\mathcal{O}}\left(n\right),\quad\tilde{J}_{x}^{s}=-\frac{2\lambda}{\epsilon}v+\mathcal{\mathcal{O}}\left(n\right)\,,\quad\tilde{J}_{x}^{sh,m}=0\,,
\end{align}
while inside the Rashba pseudogap, $\epsilon<2\lambda$, we find 
\begin{align}
\tilde{J}_{x}^{c} & =v\frac{2\left(\epsilon^{2}+\lambda\epsilon+2\lambda^{2}\right)}{\epsilon^{2}+4\lambda^{2}}+\mathcal{\mathcal{O}}\left(n\right)\,,\\
\tilde{J}_{x}^{s} & =-v\frac{\epsilon(\epsilon+2\lambda)}{\epsilon^{2}+4\lambda^{2}}+\mathcal{\mathcal{O}}\left(n\right)\,,\\
\tilde{J}_{x}^{sh} & =\frac{n\,u_{0}^{2}}{16\,v}\frac{\epsilon}{\lambda}+\mathcal{\mathcal{O}}\left(n^{2}\right)\,,\\
\tilde{J}_{x}^{m} & =\frac{n\,u_{0}^{2}}{4\,v}\frac{\epsilon^{2}}{\epsilon^{2}+4\lambda^{2}}+\mathcal{\mathcal{O}}\left(n^{2}\right)\,.
\end{align}
At leading order in $n$, the important components are $\tilde{J}_{x}^{c},\tilde{J}_{x}^{s}$.
In the strong scattering regime Eqs.\,(15)-(19) acquire logarithmic
corrections in the ultraviolet cutoff $\Lambda$. In Fig. \ref{fig:The-charge-}
we show that such corrections are small for the leading terms $\tilde{J}_{x}^{c}$
and $\tilde{J}_{x}^{s}$, so that Eqs.\,(15)-(17) still hold in this
regime. Having performed the limit $u_{0}\to\infty$, Eqs.\,(18),(19)
for the subleading components $\tilde{J}_{x}^{sh},\tilde{J}_{x}^{m}$
are no longer valid; yet we find the corrections provide a negligible
contribution to the response functions (not shown). 

\subsection{Finite temperature calculation for the ISGE efficiency}

In Fig.\,(3) of the main text we showed the figure of merit's temperature-dependence.
The calculation was performed numerically employing the following
definition 
\begin{equation}
\gamma(\mu,T)=\frac{\int_{-\infty}^{+\infty}d\epsilon\,\frac{\partial f(\epsilon,\mu,T)}{\partial\epsilon}\,2|\chi_{yx}(\epsilon,T=0)|\left[\theta(|\epsilon|-2\tilde{\lambda})v_{F}(2\tilde{\lambda})+\theta(2\tilde{\lambda}-|\epsilon|)v_{F}(\epsilon)\right]}{\int_{-\infty}^{+\infty}d\epsilon\,\frac{\partial f(\epsilon,\mu,T)}{\partial\epsilon}\,\sigma_{xx}(\epsilon,T=0)}\,,\label{eq:gamma}
\end{equation}
where $f(\epsilon,\mu,T)=\left\{ 1+\exp\left[(\epsilon-\mu)/k_{B}T\right]\right\} ^{-1}$
is the Fermi\textendash Dirac distribution function; see main text
for remaining definitions. 

\section{EFFECT OF RANDOM SOC}

\begin{figure}
\centering{}\includegraphics[scale=0.5]{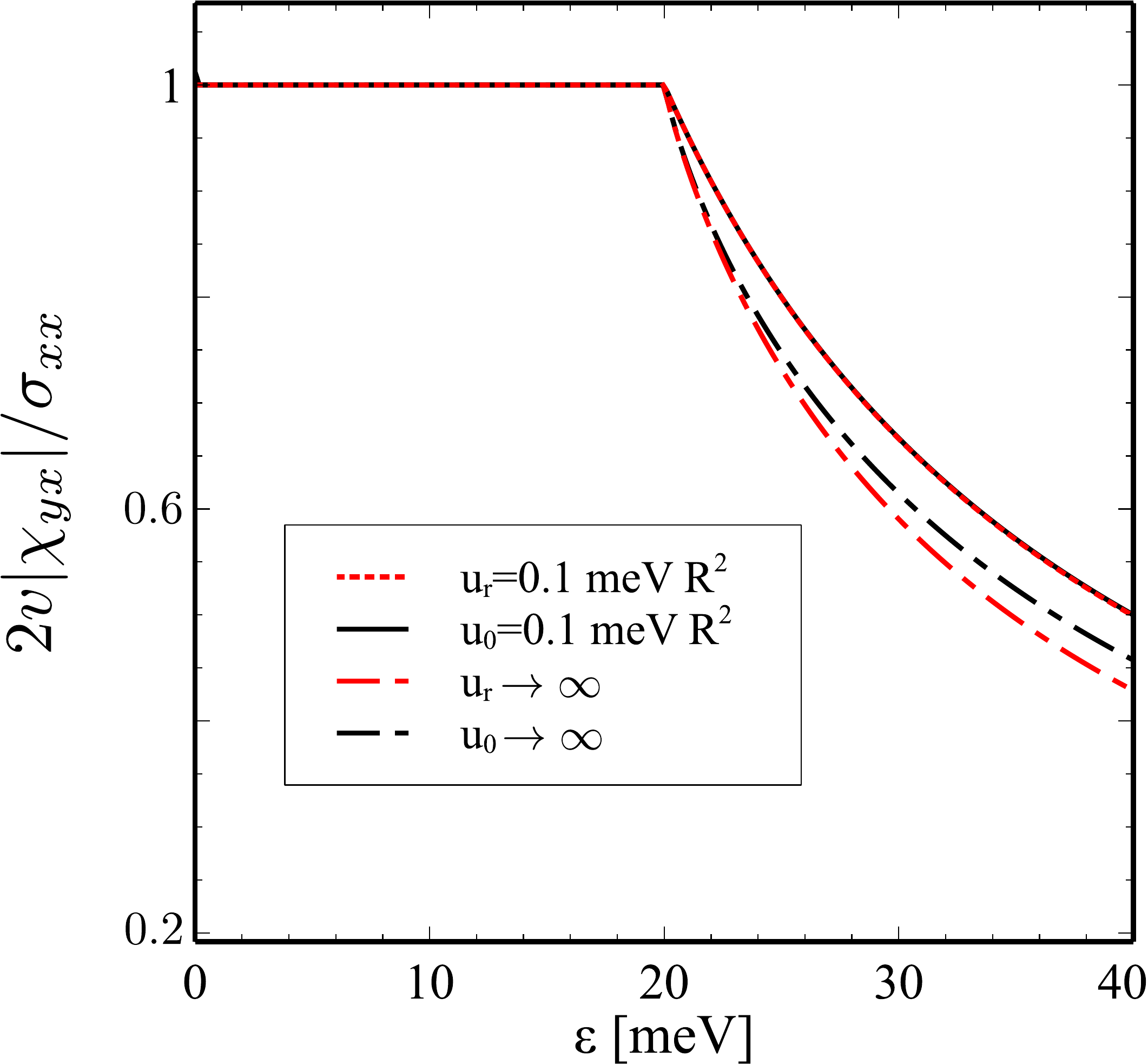}\caption{Comparison between pure scalar ($u_{0}$) and pure Rashba disorder
($u_{r}$). $R=6$ nm, $\lambda=10$ meV. While for in the weak scattering
limit the results coincide, Rashba impurities have a slightly more
important impact in the UL. However the difference between the two
cases is clearly very tiny.\label{fig:-ratio:-comparison}}
\end{figure}

We analyze here the impact of \emph{random} Rashba fields (RRFs) on
the CSC efficiency. In graphene without proximity SOC, RRFs lead to
current-driven spin polarization via asymmetric spin precession \cite{Huang_16}.
In graphene on TMD, small fluctuations in the Rashba-Bychkov coupling
($|\lambda(\mathbf{x})-\lambda|\ll\lambda$) cannot disturb the spin
helicity of eigenstates. This directly implies that the CSC rate in
regime I remains unaffected (see main text). To investigate the impact
of random SOC in regime II, we model the RRF as a short-range disorder
potential with Rashba-Bychkov matrix structure:
\begin{align}
V_{\textrm{RRF}}\left(\mathbf{x}\right) & =u_{r}\gamma_{r}\,\sum_{i=1}^{N}\delta\left(\mathbf{x}-\mathbf{x}_{i}\right)\,,
\end{align}
Neglecting its real part real, the self-energy $\Sigma^{a}$ preserves
the structure of Eq.\,10 of the main text
\begin{align}
\Sigma^{\pm} & =\mp in\left(\eta_{0}\gamma_{0}+\eta_{3}\,\gamma_{\textrm{KM}}+\eta_{r}\,\gamma_{r}\right)\,,\label{eq:self-en-1}
\end{align}
where $\gamma_{0}=\tau_{0}\sigma_{0}s_{0}$ (identity), $\gamma_{r}=\tau_{z}\left(\boldsymbol{\sigma}\times\mathbf{s}\right)\cdot\hat{z}$,
$\gamma_{\textrm{KM}}=\tau_{0}\sigma_{z}s_{z}$. We report the the
weak scattering limit form of the parameters appearing in Eq.\,\eqref{eq:self-en-1}
for positive energies
\begin{equation}
\begin{cases}
\eta_{0}=-\eta_{r}=\frac{u_{r}^{2}}{2v^{2}}\epsilon\,,\,\eta_{r}=0\,, & \epsilon>2\lambda\\
\eta_{0}=-\eta_{r}=-\eta_{3}=\frac{u_{r}^{2}}{4v^{2}}\epsilon\,, & \epsilon<2\lambda
\end{cases}\,.\label{eq:rashbaSE}
\end{equation}
We find that at leading order in the impurity areal density, Eq.\,(17)
of the main text still holds with a slightly different functional
form for $g\left(u_{r},\epsilon\right).$ In Fig.\,\eqref{fig:-ratio:-comparison}
we plot the ratio $|\chi_{yx}|/\sigma_{xx}$ for Rashba-like and scalar
impurities (as considered in the main text); the CSC ratio in the
two cases is virtually identical in the Born scattering regime. This
confirms that as long as the proximity effect is well developed in
the band structure of graphene, the CSC mechanism is robust against
random fluctuations in the energy scales of the model.

\begin{thebibliography}{10}
\bibitem{note_symmetries}Invariance under mirror reflection about
$\hat{x}$-axis and isospin rotations, $\Lambda_{z}=\tau_{z}$, reduces
Eq.~(\ref{eq:BS_equation}) to a set of $8\times8$ coupled equations.
In addition, the minimal Dirac\textendash Rashba model is invariant
under a rotation of $\pi$ exchanging sublattices, $C_{2}$, and $\Lambda_{x,y}=\tau_{x,y}\sigma_{z}$,
leading to only four allowed components.
\end{thebibliography}
\end{document}